\documentclass{kapproc} 
\usepackage{t1enc}
\usepackage{procps}
\usepackage{amssymb}
\usepackage[dvips]{graphicx}

\kluwerbib

\def\kms{\relax \ifmmode {\,\mbox{km\,s}}^{-1}\else \,\mbox{km\,s}$^{-1}$\fi}
\def\ha{\relax \ifmmode {\mbox H}\alpha\else H$\alpha$\fi}
\def\hb{\relax \ifmmode {\mbox H}\beta\else H$\beta$\fi}
\def\hi{\relax \ifmmode {\mbox H\,{\scshape i}}\else H\,{\scshape i}\fi}
\def\hii{\relax \ifmmode {\mbox H\,{\scshape ii}}\else H\,{\scshape ii}\fi}
\def\oiii{\relax \ifmmode {\mbox O\,{\scshape iii}}\else O\,{\scshape iii}\fi}
\def\oii{\relax \ifmmode {\mbox O\,{\scshape ii}}\else O\,{\scshape ii}\fi}
\def\oi{\relax \ifmmode {\mbox O\,{\scshape i}}\else O\,{\scshape i}\fi}
\def\nii{\relax \ifmmode {\mbox N\,{\scshape ii}}\else N\,{\scshape ii}\fi}
\def\sii{\relax \ifmmode {\mbox S\,{\scshape ii}}\else S\,{\scshape ii}\fi}
\def\lha{\relax \ifmmode \mbox {L}_{H\alpha}\else $\mbox{L}_{H\alpha}$\fi}
\def\ldig{\relax \ifmmode {\mbox L}_{DIG}\else ${\mbox L}_{DIG}$\fi}
\def\ls{\relax \ifmmode {\mbox L}_{ Str}\else ${\mbox L}_{ Str}$\fi}
\def\eme{\relax \ifmmode {\,\mbox{pc\,cm}}^{-6}\else \,pc\,cm$^{-6}$\fi}
\def\l{\relax \ifmmode  \lambda\else $\lambda$\fi}

\def\Msun{M$_\odot$}

\def\me{$^{-1}$}

\def\arcsec{\hbox{$^{\prime\prime}$}}

\begin{document}

\articletitle[Gas flows, star formation and galaxy evolution]
{Gas flows, star formation and galaxy evolution}
 
\author{John E. Beckman\altaffilmark{1,2}, Emilio Casuso\altaffilmark{1},
Almudena Zurita\altaffilmark{3} and M\'onica Rela\~no\altaffilmark{1}} 
 
\affil{\altaffilmark{1}Instituto de Astrof\'\i sica de Canarias, Spain,\\
\altaffilmark{2}Consejo Superior de Investigaciones Cient\'\i ficas, Spain\\
\altaffilmark{3}Universidad de Granada, Dept. de F\'\i sica Te\'orica y del Cosmos, Granada, Spain}

\begin{abstract}			
In the first part of this article we show how observations of the 
chemical evolution of the Galaxy: G- and K--dwarf numbers as 
functions of metallicity, and abundances of the light elements, 
D, Li, Be and B, in both stars and the interstellar medium (ISM),
lead to the conclusion that metal poor \hi\ gas has been accreting
to the Galactic disc during the whole of its lifetime, and is 
accreting today at a measurable rate, $\sim2$\Msun\ per year across the 
full disc. Estimates of the local star formation rate (SFR) using
methods based on stellar activity, support this picture. The best
fits to all these data are for models where the accretion rate is
constant, or slowly rising with epoch. We explain here how this 
conclusion, for a galaxy in a small bound group, is not in conflict 
with graphs such as the Madau plot, which show that the universal 
SFR has declined steadily from $z=1$ to the present day. We also 
show that a model in which disc galaxies in general evolve by accreting 
major clouds of low metallicity gas from their surroundings can 
explain many observations, notably that the SFR for whole galaxies 
tends to show obvious variability, and fractionally more for early 
than for late types, and yields lower dark to baryonic matter ratios 
for large disc galaxies than for dwarfs. In the second part of the
article we use NGC~1530 as a template object, showing from 
Fabry--P\'erot observations of its \ha\ emission how strong shear
in this strongly barred galaxy acts to inhibit star formation,
while compression acts to stimulate it.  
\end{abstract}


\section{Introduction}			
The role of mergers in galaxy evolution has become 
increasingly recognized recently, stimulated by the central role of CDM  
cosmology (Navarro, Frenk \& White 1994, 1995; Power et al. 2003). The 
importance of mergers was realized from purely observational arguments by 
Toomre \& Toomre (1972); they argued that as tidal encounters generate short
lived features, to yield today's ``peculiar'' galaxies a population of binary
galaxies with highly eccentric orbits is required. If these have a flat binding
energy distribution, their merger rate must have declined with time as $t^{-5/3}$
so that the ten obviously merging objects in the New General Catalogue must be the tail end of 
750 remnants (see also Toomre, 1977). Zepf \& Koo (1989), prior to the Hubble
Deep Fields, and Abraham (1999) using their contents, inferred that galaxy 
pair density grows as $(1+z)^3(\pm1)$, while Brinchmann et al. (1998) showed 
that irregular galaxies form 10\% of the total at $z\sim0.4$ and 30\% at $z\sim0.8$.
Peculiar morphologies, characteristic of mergers and interactions dominate
at high redshift. The route to ellipticals via major mergers of spirals was first
explored by Toomre \& Toomre (1972), by Toomre (1977), and by Verraraghavan
\& White (1985).

Although most studies focused on the more spectacular major mergers,
studies of the Galaxy have brought out the importance of minor mergers and 
accretions. For example Gilmore \& Feltzing (2000) in a major review, using 
arguments based on metallicity and kinematics, discuss its structural components:
bulge, thin disc, thick disc (Freeman 1993; Gilmore \& Wyse  1998), and 
stellar halo, showing that the thin disc and bulge are relatively young,
while the thick disc and halo are relatively old. In particular the thick disc
formed at a specific epoch, some 10 Gyr ago (Gilmore \& Wyse 1998) as 
a result of a minor merger, and the Galaxy is accreting material now as it 
disrupts its smaller neighbours (Lynden--Bell 1976; Mirabel 1982; Savage
et al. 2000). Here we will point up a process which is playing a key role
in the evolution of the Galaxy, but which tends to be overlooked: the accretion 
of gas clouds of sub--galactic mass. This must be important in galaxy groups,
small clusters, and the outskirts of rich clusters in general. We will show 
here, as the main thrust of the article, that the Galactic  accretion rate has 
not declined during the disc lifetime and while this may seem surprising, it is 
a natural consequence of a plausible model for the Local Group. To maintain the 
star formation rates (SFR's) observed in late type galaxies now a steady inflow
of gas from outside the galaxy is maintained. A key difference between a gas--rich
and a gas--poor galaxy is that the capture cross--section for gas cloud accretion
is higher in the former; this alone could account for the lower mean SFR's seen 
in late type galaxies but also for their greater relative scatter. The 
distribution of star formation across a galaxy disc is also affected strongly by 
gas flows, but here by in--plane flows. We bring this out in a semi--quantitative 
way in the final part of the paper, where we present observations of the 
relation between the velocity gradient of the gas flow in a strongly barred 
galaxy, NGC~1530, and the local formation rate of massive stars, inferring that 
compression enhances the local SFR, while shear reduces it.

\section{Direct and indirect evidence for continuous gaseous inflow to the Galactic disc}
Long before galaxy mergers were considered important, Larson (1972a,b)
suggested that long term infall of low metallicity gas to the Galactic disc
could explain the observed relative dearth of metal poor stars (the G--dwarf
problem). An advantage of this scenario over its rivals is that it offers an 
explanation for the presence of measurable deuterium abundance in the Galaxy,
notably near the centre (Audouze et al. 1976, and more recently Lubowich et al.
 2000). With no known source of D within the Galaxy, and as astration destroys 
it, continuous replenishment by infall can explain its presence. High redshift
D abundances from the Ly$\alpha$ forest (e.g. Kirkman et al. 2000; Pettini \& 
Bowen 2001) are $\sim$2-4$\times$10$^{-5}$, while recent FUSE values for the local 
ISM (e.g. Lehner et al. 2002; Oliveira et al. 2003) are $\sim$1-2$\times$10$^{-5}$, 
and to complete the argument for infall the D abundance recently measured with FUSE in the high
velocity cloud (HVC) ``complex C'' (Sembach et al. 2004) is 2.2$\times$10$^{-5}$, a value 
intermediate between the first two. The idea is that some HVC's are supplying 
metal--poor non--astrated gas continuously to the Galaxy: they are the infall 
component mentioned. Some early models of Galactic inflow (Hunt \& Sciama,
1972; Tinsley 1977), which derived infall mass rates similar to the Galaxy--wide SFR, 
entailed a sweeping up of Local Group gas, while Larson (1976) 
suggested collapsing remnants of the pregalactic nebula. Muller, Oort \& Raimond
(1963) detected HVC's in \hi\ at 21 cm; Oort (1970) noted that some of them
have radial velocities higher than the local Galactic escape velocity, and
might be primordial. Recent observations that their metallicities are $\sim0.1$ solar
(e.g. Lehner et al. 2002; Wakker et al. 2003; Savage et al. 2003; Sembach et
al. 2004), rule out primordiality, but also rule out a Galactic origin, so 
they are candidates for low metallicity infall. Evidence that some HVC's may 
well belong to the local group but not the Galaxy had been accumulating (Mirabel 
1982; Lepine \& Duvert 1994; Blitz et al. 1999; L\'opez--Corredoira et al. 
1999; Casuso \& Beckman 2001; Gibson et al. 2001; Putman et al. 2002;
and recently Braun et al. 2002) in a sensitive \hi\ survey, estimated that 
a gas mass of a few times 10$^9$\Msun\ could have accreted to both the Milky Way and M31
during their lifetimes. Direct evidence for infall for other galaxies is small.
Phookun et al. (1993) detected \hi\ falling onto NGC~5254, while recently 
Fraternali et al. (2004) detected flows perpendicular to the planes of nearby spirals
using \ha\ as the tracer. Observations of infall to external galaxies are few
so far because the detection sensitivity has been limited at $\sim$10$^8$\Msun\ of gas,
while it is probable that the bulk of these flows are in clouds of lower mass.
As sensitivities have improved, signs of accretion are being reported, see e.g.
van der Hulst \& Sancisi (2003).
\begin{figure}[ht]
\centerline{\includegraphics[width=5cm]{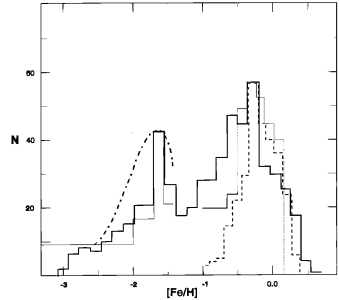}}
\vspace{-0.3cm}
\caption{Early frequency histogram of local G dwarfs as a function of 
metallicity (using [Fe/H] as parameter) from CB97. Two peaks, for halo and 
thin disc stars, are clearly seen. The thick disc contribution, less obvious, 
lies between them  and widens the higher metallicity peak (for the disc stars).}
\label{fig1}
\end{figure}

The G--dwarf problem is a strong pointer to infall for the Milky Way
(van den Bergh 1962; Schmidt 1963; Pagel \& Pachett 1975; Tinsley 1980; 
and Pagel 1987). Closed box models for chemical evolution of the Galactic 
disc predict far more low metallicity stars than those observed, and integrated
population studies of external galaxies point to the same conclusion (Worthey
et al. 1996; Espana \& Worthey 2002; Bellazzini et al. 2003). Late F and
G stars may be nearly as old as the disc, and preserve a record of its evolution
though the oldest have evolved off the main sequence, so as improving technique
brought K--dwarf statistics within range (Flynn \& Morrell  1997) their local 
metallicity distribution has given an more accurate history to follow. Excluding
halo and thick disc stars kinematically, a frequency $v.$ metallicity plot
for G, or better K dwarfs show low numbers for low disc metallicity ([Fe/H]$<-1$) 
rising sharply to a narrow peak between [Fe/H]=-0.6 and -0.2, then
falling off to higher metallicities. The earlier G dwarf statistics (Carney 
et al. 1990; Rocha--Pinto \& Maciel 1996) are well explained in a scenario
with near constant infall of low metallicity gas to the Galactic plane (Casuso
\& Beckman 1997). The K dwarfs show the same metallicity pattern (Flynn \&
Morrell 1997; Favata et al. 1997; Kotoneva et al. 2002) and we will discuss
this and how it is modelled in Sect.~4 below. Fig.~1 shows an early stellar
metallicity--frequency plot, taking in thin disc, thick disc and halo G--dwarfs
from Casuso \& Beckman (1997) (hereinafter CB97).

\begin{figure}[ht]
\centerline{\includegraphics[width=5.4cm]{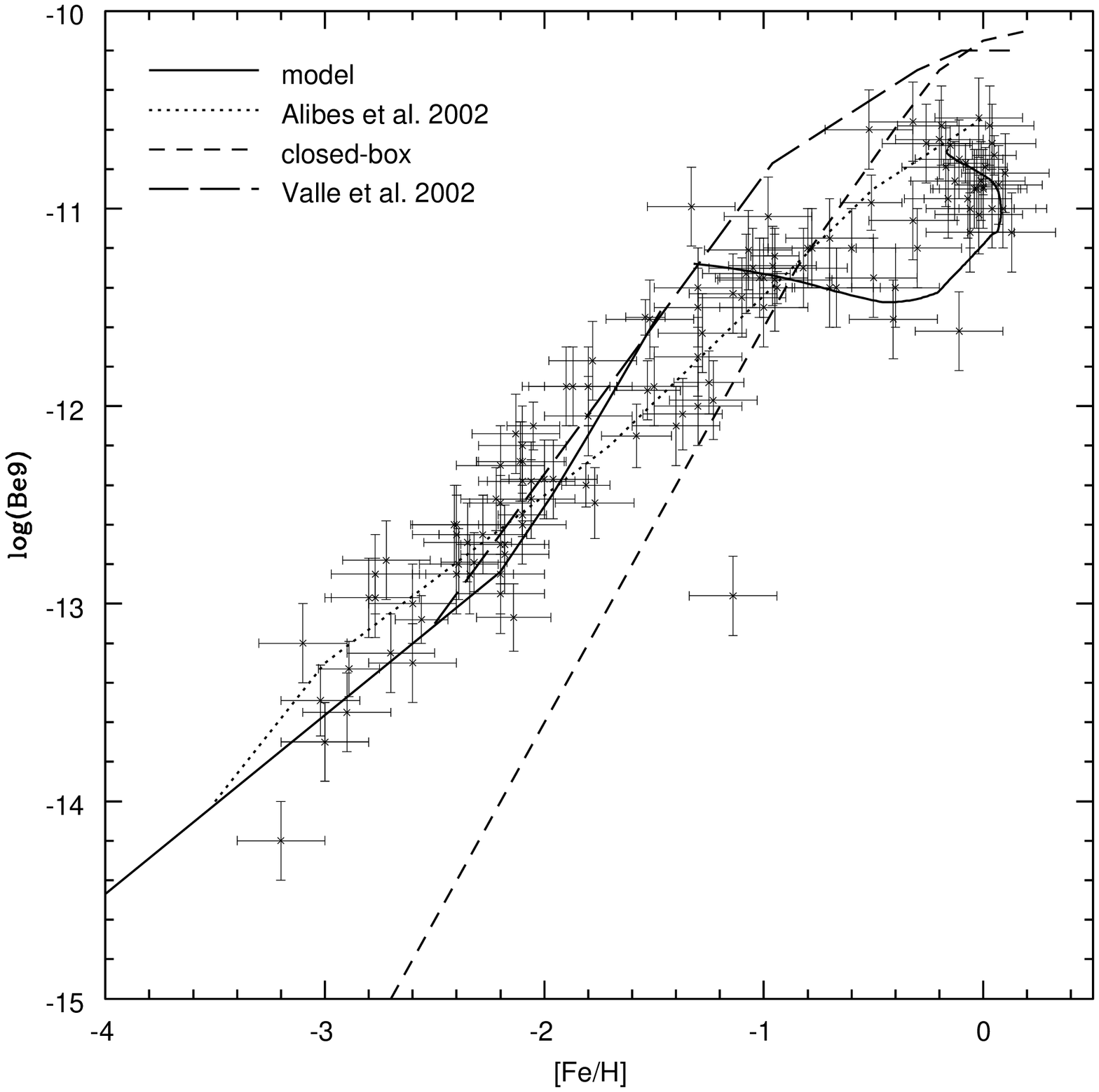}
            \includegraphics[width=5.4cm]{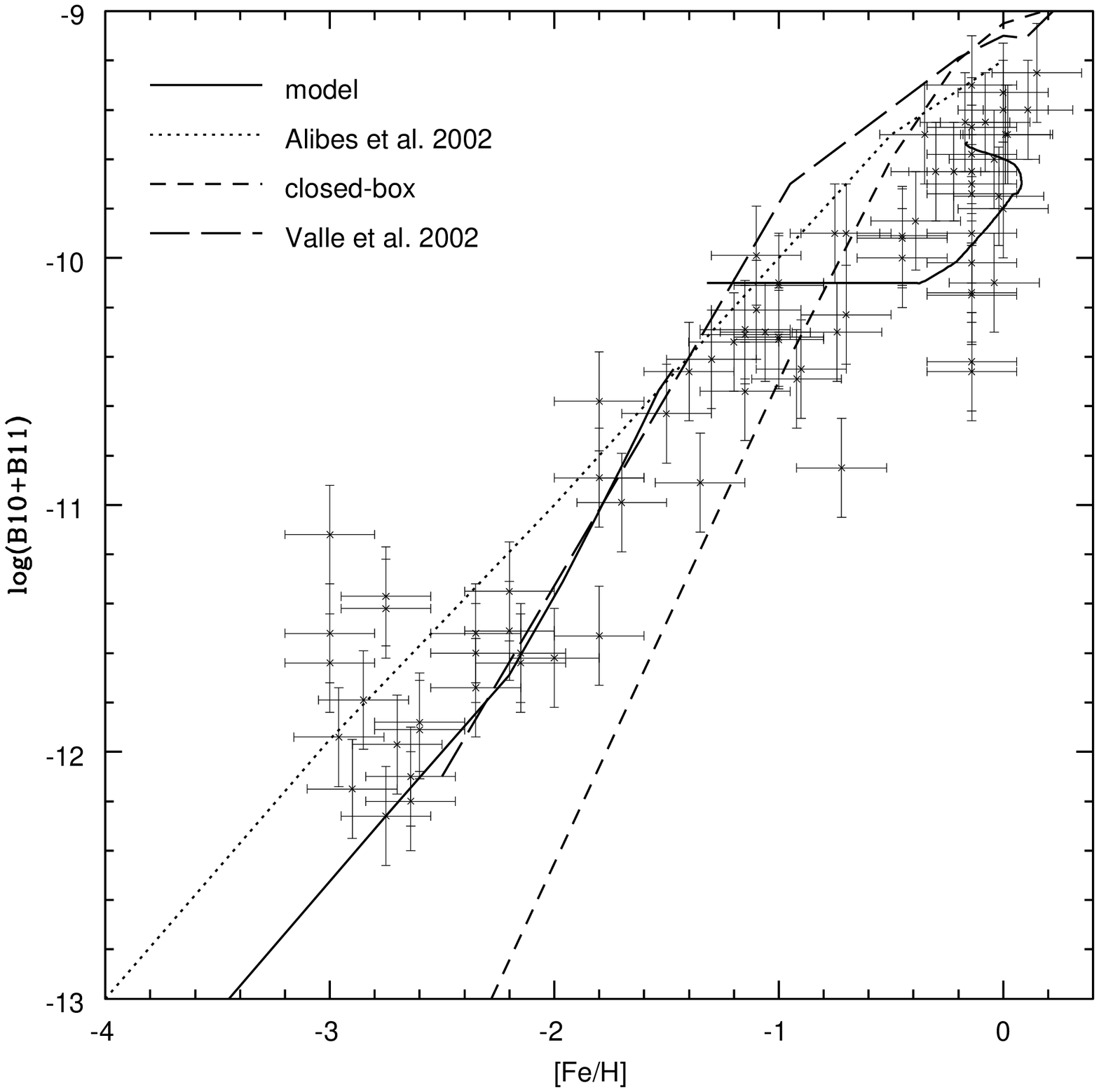}}
\vspace{-0.5cm}\caption{{\bf (left)} Compilation of Be abundances measured in local disc 
and halo 
stars (taken from CB04). Model predictions shown are for a closed box model,
for two models with infalling \hi\ but at rates which decline with time
(Alib\'es et al. 2002; Valle et al. 2002) and one model, for the disc,
with slowly rising infall: CB97, solid line. For more details see text.
{\bf (right)} Compilation of B abundances measured in local disc and halo stars,
taken from CB04. Model predictions are from the same sources as in Fig.~2 (left).}
\label{fig2}
\end{figure}

	Further chemical evidence explained well by low metallicity infall is 
the evolution of beryllium and boron, measured $v.$ iron as standard in local 
disc stars. A recent compilation of observational data on this is shown in 
Fig.~2 (left and right). The ``loop--back'' seen for disc stars in both plots is well 
explained by infall, and best explained if the infall rises slowly with time.
Recent models for Be and B evolution in the Galaxy (Fields et al 2000;
Ramaty et al. 2000) have focused on their evolution in the halo, i.e. for 
[Fe/H]$<-1$, invoking specific mechanisms for their production by high cosmic
ray fluxes in star forming zones to explain the observed linear relations of 
both Be and B with Fe. But their treatment of the disc evolution of Be and B
lacks finesse, notably in the chemical evolution model used, mainly  because
the authors are not really aiming to explain the disc observations, even though
over 90\% of Be and B were made in the disc. Alib\'es et al. (2002) use a much 
better disc model, but even they do not predict very well the observations 
in Fig.~2. In CB97 we had managed to model earlier versions
of these data sets. To our surprise, even concern, the model entailed \hi\ infall 
with a rising rate over the disc lifetime. In 1997 the B observations were sparse
but as shown in Fig.~2 the trends for B and Be $v.$ Fe have been strongly confirmed
as data has accumulated. It was these B and Be observations which first pushed
us towards an increasing inflow rate scenario, but other types of evidence have
accumulated to support this as we now describe.

\begin{figure}[ht]
\centerline{\includegraphics[width=7cm]{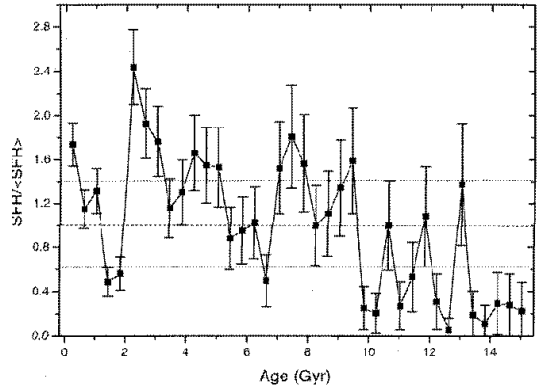}}
\caption{Star formation rate as a function of epoch, from Rocha--Pinto et al. (2000)
based on stellar activity, and from observations of stars within 1~kpc of 
the Sun. Although absolute ages are progressively less accurate as they increase,
the general run of SFR has clearly increased with increasing time, on which major
shorter term variations are superposed.}
\label{fig3}
\end{figure}

\section{The history of the local star formation rate (SFR)}
Since Vaughan \& Preston (1980) brought out the possible use of 
chromospheric activity in late type dwarfs as a chronometer, a number of surveys
within 1~kpc of the Sun have used activity indices to investigate the time dependence
of the local SFR. Barry (1988) found using this technique that the SFR, averaged
in bins of 10$^9$yr showed evidence of a secular increase during the disc 
lifetime, but also large amplitude excursions in this general trend. From a much
wider stellar sample Rocha--Pinto et al. (2000) confirmed both findings, as we 
show here in Fig.~3. Although both samples were taken locally to 
the Sun, Wielen (1977, 1996) showed that the diffusion of stellar orbits implies
that such a sample must include stars which have drifted radially from their 
birthplaces by over 2~kpc, both inwards and outwards. Thus any large excursion 
in the SFR must have occurred simultaneously over a major fraction of the 
Galactic disc. Interaction could cause this, but not only with another galaxy; 
accretion of a large gas cloud will also work. The data in Fig.~3 are well 
interpreted in a scenario of \hi\ infall of clouds with a range of sizes. We can 
estimate a time scale for single cloud accretion, rough as it depends on 
distance estimates, using measured properties of extragalactic HVC's. From
Blitz et al. (1999) or Putman et al. (2002) we set a lowish scale size for a 
large cloud at $\sim$5~kpc. Measured radial velocity of $\sim$200~km~s$^{-1}$ 
then gives an  accretion time of a few times 10$^8$~yr, which is consistent with the widths of 
the surge peaks in Fig.~3. Also from Fig.~3 we can see a long term slowly rising
trend, which is clearly worth further comment.

	Is this apparently rising SFR compatible with cosmological trends?
Since the Madau et al. (1996) plot it is known that the global SFR in the 
universe has been falling steadily from $z=1$ to the present epoch. We must note
though that the Local Group is gravitationally bound; interaction rates between
galaxies have been steady since the group condensed. The HVC accretion rate 
to the Milky Way must depend on the density of the clouds within the Local 
Group, the gravitational range of the Galaxy, and its accretion cross--section.
While accretion causes the first parameter to fall with time, it causes the
second and third to rise as gas accumulates in the plane. In L\'opez--Corredoira
et al. (1999) we computed that for the current accretion rate to be rising (or
constant in the limiting case) the present mean cloud density must have a lower
limit, which can be converted to an estimate of the fractional mass of the 
Local Group in the form of these intracluster clouds. The value is $\sim$50\%, which
is fully plausible, comparing the masses of its galaxies (the Milky Way and M31
as the smaller objects contribute little) with dynamical estimates of the
Local Group mass, as first done by Kahn \& Woltjer (1959); for a recent estimate
see Whiting (1999). (Note that these conclusions depend on only one assumption 
about dark matter: that the baryon to dark matter density does not vary much
from object to object, which we cannot discuss in detail here, but we take as
reasonable). We have also shown, in L\'opez--Corredoira et al. (2002) that 
if the net long term vector of this inflow is directed to the Local Group 
barycentre (Blitz et al. 1999) its dynamic parameters are those required
to yield the known Milky Way \hi\ warp amplitude and direction. A global check 
on the inflow rate can come from measurements of the SFR. For example SN II rates,
of order a few per century in the disc, (e.g. Dragicevich et al. 1999) agree
with an accretion rate of a few \Msun~yr$^{-1}$, using a Salpeter IMF with mass limits
0.1\Msun\ and 100\Msun, and that type II SNe come from stars with M$>$8\Msun. Chemical
evolution models require mean accretion rates during the disc lifetime of 
$\sim$2\Msun~yr$^{-1}$, consistent with estimates of HVC accretion rates (Blitz et al. 1999;
Braun et al. 2002).

\begin{figure}[ht]
\centerline{\includegraphics[width=6cm]{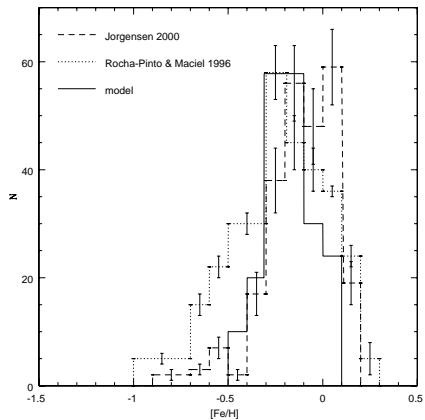}}
\vspace{-0.5cm}
\caption{Frequency histograms of metallicity for G dwarfs in the local Galactic
disc, from Rocha--Pinto \& Maciel (1996), and from Jorgensen (2001). In the 
latter a scale height criterion has been used to minimize the thick disc 
contribution. Compared with these data is a model based on chemical evolution 
with increasing \hi\ infall from CB04.}
\label{fig4}
\end{figure}

\section{The K-dwarf problem}
	The dearth of low metallicity K dwarfs in the Galactic disc is a more
severe test of chemical evolution models than previous similar data for G 
dwarfs. In a recent article (Casuso \& Beckman 2004) we describe the
observational and theoretical developments in this field and this section gives 
an overview of these. The G--dwarf data by Rocha--Pinto \& Maciel (1996) already
showed a much narrower peak in the metallicity frequency distribution than 
that in Fig.~1. Using a chemical evolution model (CB97) developed to explain the 
Be and B data, in Casuso \& Beckman (2001) (CB01) we modelled this narrow peak 
well. The model included a rising infall rate, and predicted even better the
G--dwarf data of Jorgensen (2001), published while CB01 was submitted, so this
was a genuine prediction and not a post--diction as one so ofter finds. It is 
significant that we selected the best model for its Be, B predictions, although
improvements were possible using only the Rocha--Pinto \& Maciel (1996) data, so the
improved agreement with Jorgensen was remarkable. One feature of Jorgensen's 
data was the careful exclusion of thick disc stars from the sample. In Fig.~4 
we compare the two data sets with the model from CB01; these fits,especially to
Jorgensen's data, strengthened our ``increasing infall rate'' hypothesis.The next
step was to use K-dwarf metallicity distributions. In Fig.5 we show two data 
sets, by Favata et al. (1997) and Kotoneva et al. (2002), compared with 
our model prediction.We can see that agreement is good in the metallicity range
close to solar, but less good outside this range. Of the two data sets,one
lies above and the other below our prediction, at lower metallicities. The 
cause in both cases is the thick disc population. Favata et al. (1997)  cut off their
graph where the thick disc numbers become appreciable, while Kotoneva et al. (2002)
did not take special steps to exclude thick disc stars. We conclude that the
K--dwarf metallicity distribution in the thin disc gives some support to our
model, but further measurement and analysis are needed here.
\begin{figure}[ht]
\centerline{\includegraphics[width=6cm]{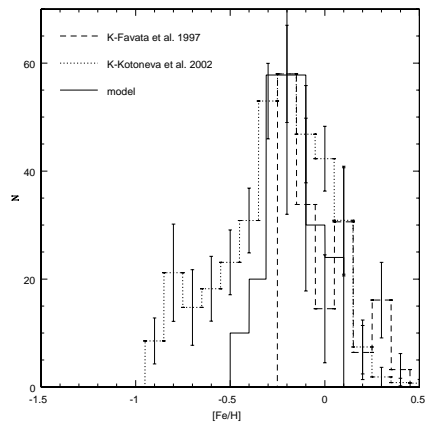}}
\vspace{-0.5cm}
\caption{Frequency histograms of metallicity for K dwarfs in the local Galactic 
disc, from Favata et al. (1997) and Kotoneva et al. (2002). Compared with 
these data we show a model based on chemical evolution with increasing 
\hi\ infall from CB04. Agreement is not as good as for the G--dwarfs in Fig.~4,
but no attempt to limit the sample to thin disc stars was made in either
observational paper, while the model is strictly for thin disc evolution.}
\label{fig4}
\end{figure}

\section{The light element abundances in the Galactic Disc}
The light elements: hydrogen, helium, lithium, beryllium
and boron, are exceptional in that their main production sites are principally
non-stellar. Hydrogen (H, D and T)  most present day helium ($^4$He and $^3$He), and 
some 10\% of lithium are primordial, while the remaining lithium, beryllium and
boron were formed in the interstellar medium (ISM) in processes involving  
Galactic Cosmic Rays (GCR). Reeves (1974) reviewed light element nucleosynthesis
setting the scene for all subsequent developments. For the present article
the importance of the light elements is that they present sensitive tests of 
abundance dilution by infall of element--poor gas into the ISM where they are
being produced. We have presented the Be and B $v.$ Fe data plots in Fig.~2, and
in Fig.~6 we show a similar compilation for Li. For Be and B there has been 
much interest in why they show linear dependences on metallicity; they are 
spallation products of CNO so initial predictions were for a quadratic 
dependence, as secondary elements. Without entering in depth in this complex
field, it is clear from Fig.~2 that this linear relation holds only in 
the halo metallicity range ([Fe/H]$<$-1.5) while for disc metallicities 
the dependence is far from linear. Clearly halo and disc behaviour are different,
and this can be explained independently of the specific GCR production 
mechanisms for Be and B by differences in the surrounding gas flow conditions.
For the halo stars, not the subject of this article, a model in which the 
star forming gas is flowing inwards to the Galactic centre as the initial 
spherical collapse occurs, gives a satisfactory explanation of the pseudo linear
Be and B $v.$ Fe observations. For the disc stars we show four model comparisons,
a closed box model and three inflow models. The best fits are for secularly 
slowly increasing inflow. The same model, with no fine tuning, was used to fit
the G- and K--dwarf distributions shown in Figs.~4 and 5, and is used as a 
framework for predicting the Li $v.$ Fe plot in Fig.~6, though here the 
importance of both production and destruction mechanisms is such that the 
goodness of fit is not a simple test of the gas flow model (see Casuso \&
Beckman 2000  for more details). Finally we can return to the case
of deuterium which we discussed briefly above. As deuterium is destroyed in 
stars, there are no equivalent plots to those in Fig.~2 for this nuclide.
However the mere presence of D in the ISM of the Galaxy today, especially near
the Galactic centre, implies very recent infall, and we discussed implications 
of this in Casuso \& Beckman (1999). To summarize this section, the 
best fits to the observed evolutionary plots of light element abundances in 
the Galactic disc are provided by models in which the infall of \hi\ has occurred
at a constant or slowly rising rate.

\begin{figure}[ht]
\centerline{\includegraphics[width=5.5cm]{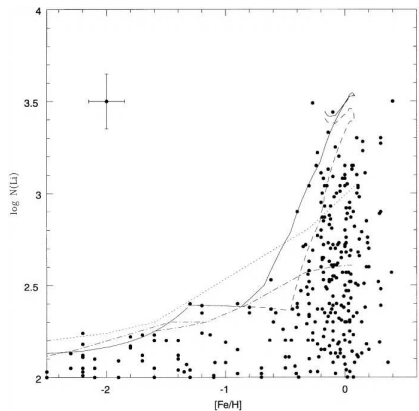}}
\vspace{-0.3cm}
\caption{Lithium abundance as a function of metallicity ([Fe/H]) for 
local disc stars; compilation from Casuso \& Beckman (2000). The upper envelope represents minimum
Li depletion in the star observed, and is the datum for model fitting. Compared 
are a number of models, also from Casuso \& Beckman (2000). The complexity of the production and 
destruction processes implies that model fits do not tell us too much about the 
time dependence of gaseous infall for Li, but all models showing fair fits to 
the upper envelope use constant or increasing infall assumptions.}
\label{fig6}
\end{figure}

\section{Gas accretion in the general process of galaxy evolution}
	Far less attention has been paid to gas accretion as a driver of 
galaxy evolution than to mergers, because the latter are more spectacular and
much easier to observe. The lower end of the galaxy mass function should 
contain clouds whose baryonic mass could be as high as 10$^8$\Msun, and would
range downwards from there. Their numbers should be high, conforming to general
laws of mass fractionation, which Elmegreen (2002) has characterized as fractal
in his discussion of Galactic gas clouds. Estimates at the low end of the 
galaxy mass function by Bell et al. (2003) confirm that the number of objects
is rising as the gas to stars ratio is rising, for the faintest dwarf galaxies,
and should keep rising for ``failed galaxies'' which are the gas clouds we ``need''.
Rosenberg \& Schneider (2002) find a slope of -1.5 for the \hi\ mass function for
galaxies. $\lambda$CDM model cosmologies (see Kauffmann, White \& Guiderdoni 
1993, for an early model including $\lambda$) do predict a steep mass function
at the low mass end, though they also overpredict satellite galaxy numbers,
which is a problem. So we would expect galaxy groups and clusters to contain 
clouds in the mass range proposed here, but most such objects lie below present \hi\ 
detectability limits ($\sim$10$^7$\Msun) and can be picked up only in nearby galaxy groups.

One result of considerable relevance is that by Kennicutt, Tamblyn 
\& Congdon (1994). In a survey they found that early type spirals have much
lower current SFRs than late types (as expected) but a much higher fractional
variability in the SFR. Tomita et al. (1996) inferred that their results imply 
SFR variability on timescales of 10$^8$ years, and Hirashita \& Kamaya (2000)
modelled this using a scenario with quasi--periodic limit cycles. One problem
with this mechanism is to explain how SFR surges can occur over a whole galaxy.
We have shown how the SFR excursions in Rocha--Pinto et al. (2000) imply Galaxy--wide
changes, which are consistent with triggering by external \hi\ clouds if these
are large enough to affect a large part of the Galaxy: sizes of several kiloparsecs
for the ``external'' HVC's were estimated by Blitz et al. (1999) and Putman et.
al (2002). The variability difference between early and late type galaxies noted 
above is because a gas rich disc presents a larger cross--section for cloud 
accretion, and a higher probability for a star formation surge when a cloud is
captured. Late types respond only to the most massive incoming clouds, and do 
so in a less dramatic way.

The current gas accretion rate of 2\Msun~yr$^{-1}$ cited in Sect.~3, 
assuming constant infall, would imply 2$\times$10$^{10}$\Msun\ of gas accreted in the disc 
lifetime; an exponentially rising rate starting from zero would give
1.4$\times$10$^{10}$\Msun. Using a dark to baryonic matter ratio of 5 the total mass accreted 
would be between 7 and 10 times 10$^{10}$\Msun. These total masses will be upper limits as 
the dark matter component of any HVC with incoming velocity higher than escape 
velocity will not be retained (this would give a natural explanation for 
possible low dark matter: baryonic matter ratios in massive late type spirals).
These predictions are just compatible with observationally based estimates.
Nikiforov et al. (2000) estimate a Galaxy disc mass of 6$\times$10$^{10}$\Msun\ within a 
20.5~kpc radius, in a total Galaxy mass of 3.3$\times$10$^{11}$\Msun. Even if these figures
are rough, we find that a major fraction of the disc has been accreted by 
gas infall. This is qualitatively compatible with ``inside--out'' evolution of
the disc, in which growing gas column density as a function of time and radius
have produced increased SFRs at smaller radii, which in turn yields a radial
metallicity gradient in the disc. Inside--out disc formation occurs in many
$\lambda$CDM based scenarios (e.g. Samland \& Gerhard 2003). In any model
where the disc is built up by gas accretion the column density threshold for 
quiescent star formation will give higher SFR's in the inner disc, although
diffusion, especially if impelled by a bar, will tend to blur out the 
radial distribution of stellar age and metallicity. As it is extremely hard, as
yet to estimate stellar population age distributions from outside a galaxy,
the local measurements we have used to predict the time dependence of infall
must be used as the best benchmark for some time to come.
\vspace{-0.4cm}
\begin{figure}[!ht]
\centerline{\includegraphics[width=4.0in]{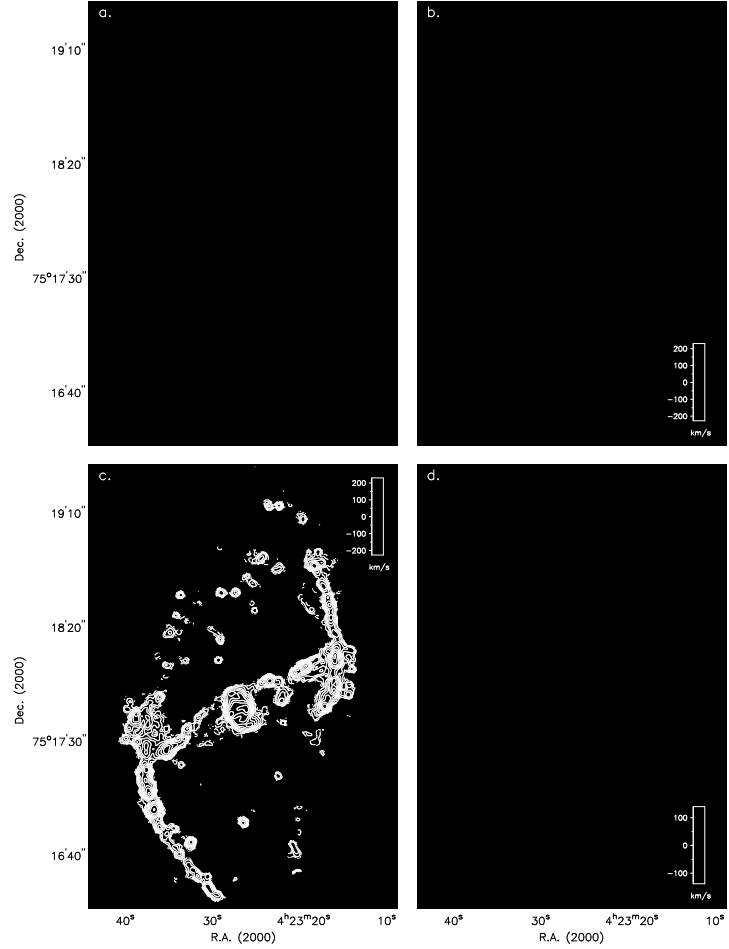}}
\vspace{-0.4cm}
\caption{{\bf(a)} Surface brightness map in \ha\ of NGC~1530, from a TAURUS data cube
from the 4.2m WHT La Palma. {\bf(b)} Radial velocity map of ionized gas from the
peaks of the \ha\ emission lines across the face of the galaxy, using the
same data cube. {\bf(c)} Contours of \ha\ surface brightness superposed on a
two--dimensional projection of the rotation curve derived from the velocity
field shown in (b). {\bf(d)} Map of the residual, non--circular velocity field,
obtained by subtracting off the projected rotation curve in (c) from the 
complete velocity field in (b). The strong non--circular velocity field aligned
with the bar is clearly seen here, as the galaxy inclination causes the 
flow along one side to be directed towards us, and away from us along the other
side of the bar (see text for more details, also see Zurita et al. 2004).}
\label{mos_gal}
\end{figure}

\section{And now for something completely different: how gas flows affect
star formation in a barred galaxy}

\subsection{Non--circular velocities around the bar}
\begin{figure}[ht]
\centerline{\includegraphics[width=3.5in]{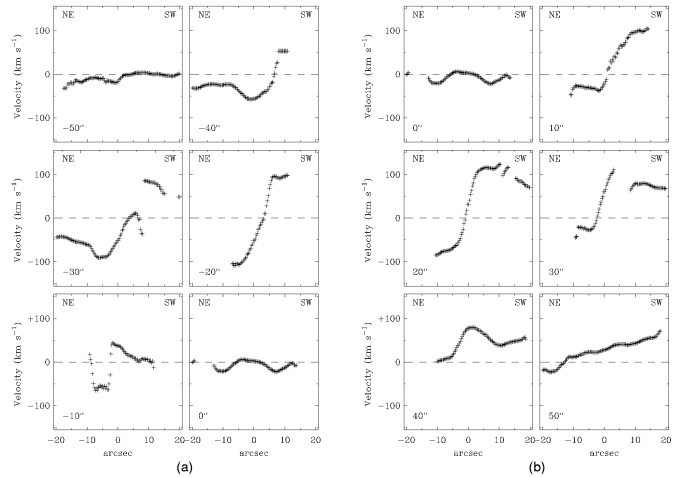}}
\caption{Cross--sections perpendicular to the bar of the non--circular
velocity field illustrated in Fig.~\ref{mos_gal}d, showing the amplitude of the
flows, $>$100~km~s\me\ in opposite directions at either side of the bar.}
\label{perfiles}
\end{figure}

So far we have been discussing accretion inflow perpendicular to the
planes of galaxies. In this shorter final section we will see how gas flows 
within the plane affect the local SFR on short timescales. For this we use
our recently measured velocity field of the strongly barred galaxy NGC~1530.
In Zurita et al. (2004) we describe the TAURUS Fabry--P\'erot interferometer on the
4.2m WHT, La Palma (Spain) with which we made the two dimensional velocity map of 
the ionized hydrogen using \ha\ emission, and give details of the observations 
and reduction. The map has a velocity interval of $\sim$18.6~km~s\me\ and an angular  
resolution of $\sim$1\arcsec, which mark improvements on previous optical and
radio maps by Regan et al. (1996), by Downes et al. (1996) and by Reynaud \&
Downes (1997, 1998, 1999). We derived first the classical rotation curve, and
used its two dimensional projection to subtract off from the observed velocity
field, yielding a residual field of non--circular projected velocity over the
full face of the galaxy. In Fig.~\ref{mos_gal}  we show how this was 
done and illustrate our results. NGC~1530 is an excellent subject for kinematic
analysis. Its arms, bar and circumnuclear disc are very well separated; the
major axis and bar position angle are well separated, and the inclination permits
good velocity and intensity measurement. In Fig.~\ref{mos_gal}c, streaming in the arms due 
to the disc density wave system shows up as ripples in the two dimensional
velocity field, with amplitude 20-30~km~s\me, which we will not analyze further 
here. In Fig.~\ref{mos_gal}b the non--circular motions round the bar show up clearly as
distorted isovels. These motions are well seen in Fig.~\ref{mos_gal}d: the residual 
non--circular velocity field, after subtracting off the rotational field seen in 
Fig.~\ref{mos_gal}c. Cross--sections through this field are shown in Fig.~\ref{perfiles},
revealing counterflow with amplitude $\gtrsim\pm$100 km~s\me\  
top either side of the bar.

\begin{figure}[ht]
\vspace{-0.3cm}
\centerline{\includegraphics[width=2.5in]{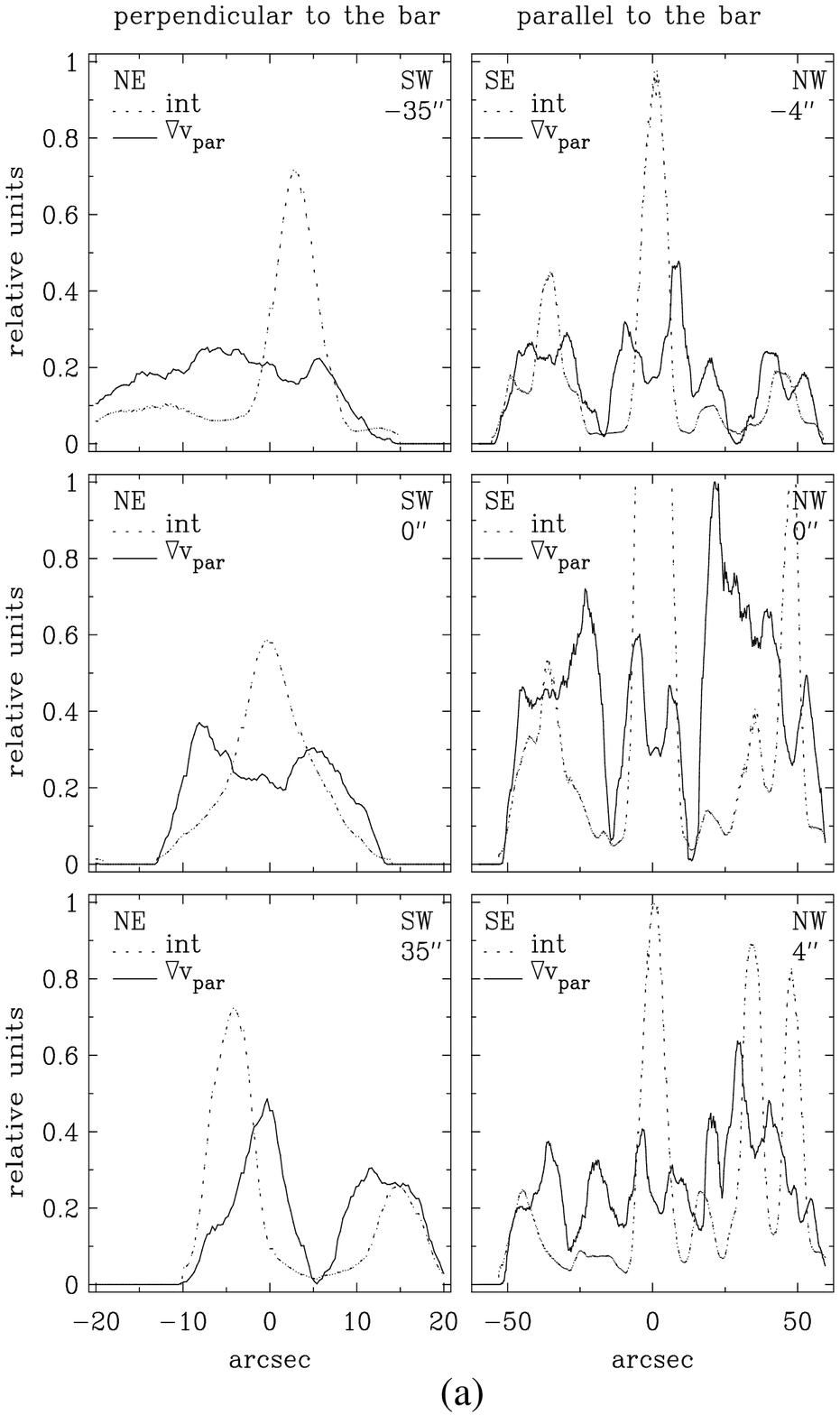}
\includegraphics[width=2.5in]{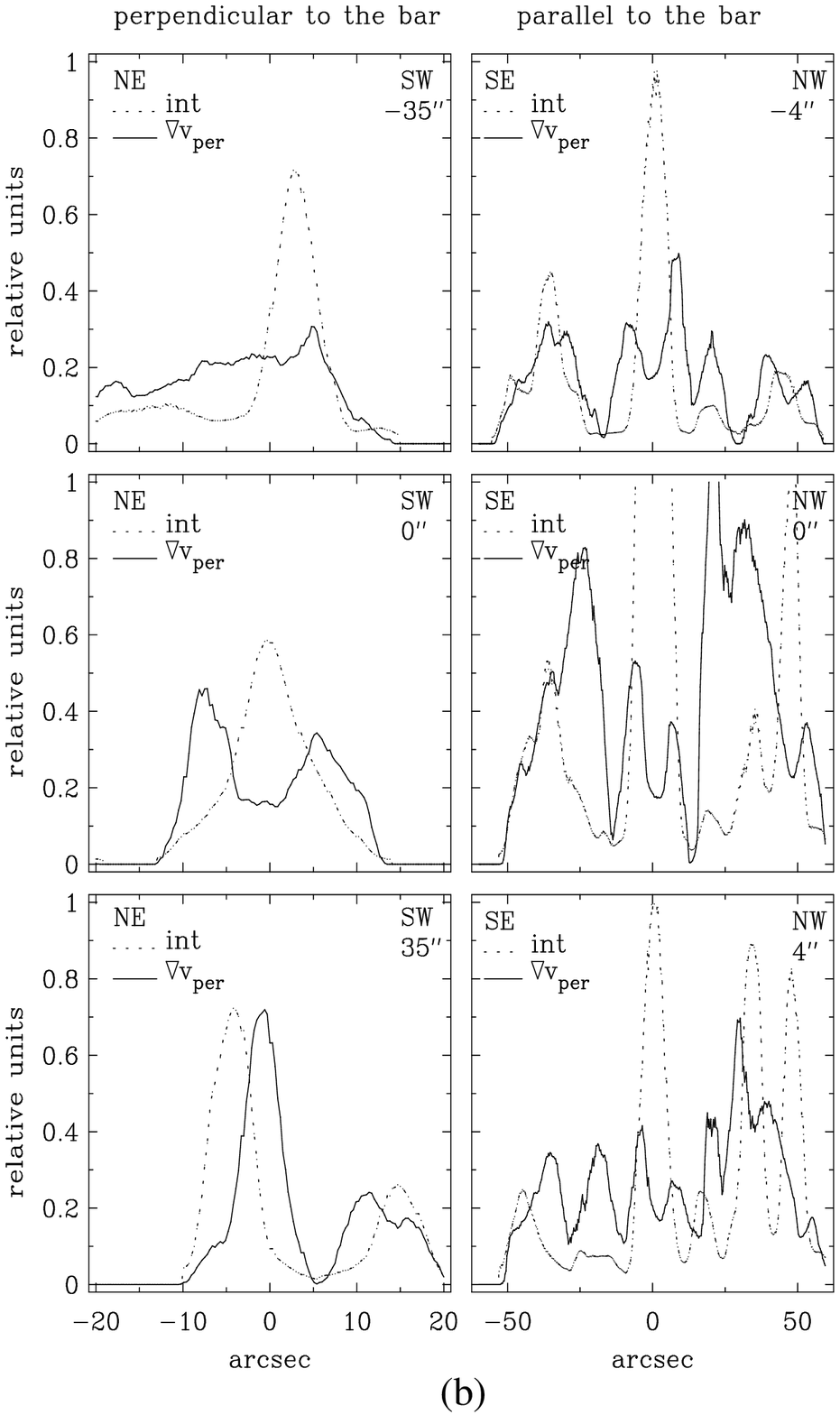}}
\vspace{-0.3cm}
\caption{{\bf (a)} Cross--sections, in the directions indicated, of the surface 
brightness in \ha\ and of the gradient in velocity parallel to the bar.
{\bf (b)} Cross--sections in the directions indicated, of the surface brightness
in \ha\ and of the gradient in velocity perpendicular to the bar. In both 
graphs we clearly see local maxima in brightness (SFR) at local minima in 
velocity gradient. For further explanation see text, also Zurita et al. (2004).}
\label{int+grad}
\end{figure}

Although \ha\ gives patchy velocity fields compared with \hi, we 
have high spatial resolution to compensate. This allows us to produce maps
of non--circular velocity gradient across the whole galaxy from a few hours'
observations. In a poster complementary to this article (Zurita et al. 2004a)
we show a beautiful result of this work: the complete coincidence of the
strong dust lanes along the bar with the loci of maximum velocity gradient
perpendicular to the bar. This shows that lines of maximum shear, which are
also lines of net zero non--circular velocity, are zones to which the dust
migrates. We can draw more powerful inferences by superposing cross--sections
of our maps in velocity gradient on cross--sections across the surface brightness
map (Fig.~\ref{mos_gal}a); the results are shown in Fig.~\ref{int+grad}. These
show clearly an anticorrelation between local SFR (\ha\ surface brightness)
and local velocity gradient,but in two distinct regimes. In zones of high
shear (gradient perpendicular to the flow along the bar) star formation is
suppressed, a conclusion inferred by Reynaud \& Downes (1998) from CO mapping.
This is as expected if shear disrupts large molecular clouds on timescales
shorter than those for massive star formation (see Kennicutt 1998 for a
discussion of this and related points). In zones of high compression star 
formation is enhanced, but occurs offset by hundreds of parsecs from the zone of 
maximum present compression. The offset relation; compression--high local SFR is
seen on scales of hundreds of parsecs. On smaller scales we see the effects of 
outflow from luminous \hii\ regions which impinges on the surrounding ISM, 
yielding ``walls'' of velocity gradient around these regions. These can be seen
in Fig.~1 of Zurita et al. (2004a), and show up here in Fig.~\ref{int+grad}, where each peak
in the \ha\ surface brightness due to a bright \hii\ region coincides with a dip
surrounded by symmetrically offset peaks, in the velocity gradient. These 
outflows have been reported by Rela\~no et al. (2003), and have amplitudes of 
40--80~km~s\me\ from the centre of the \hii\ region. We will present a quantitative
treatment of these outflows and some initial models in Rela\~no et al. (2004). 

\subsection{Gas flow spiralling to the nucleus in the inner disc}
\begin{figure}[ht]
\vspace{-0.5cm}
\centerline{\includegraphics[width=4.5in]{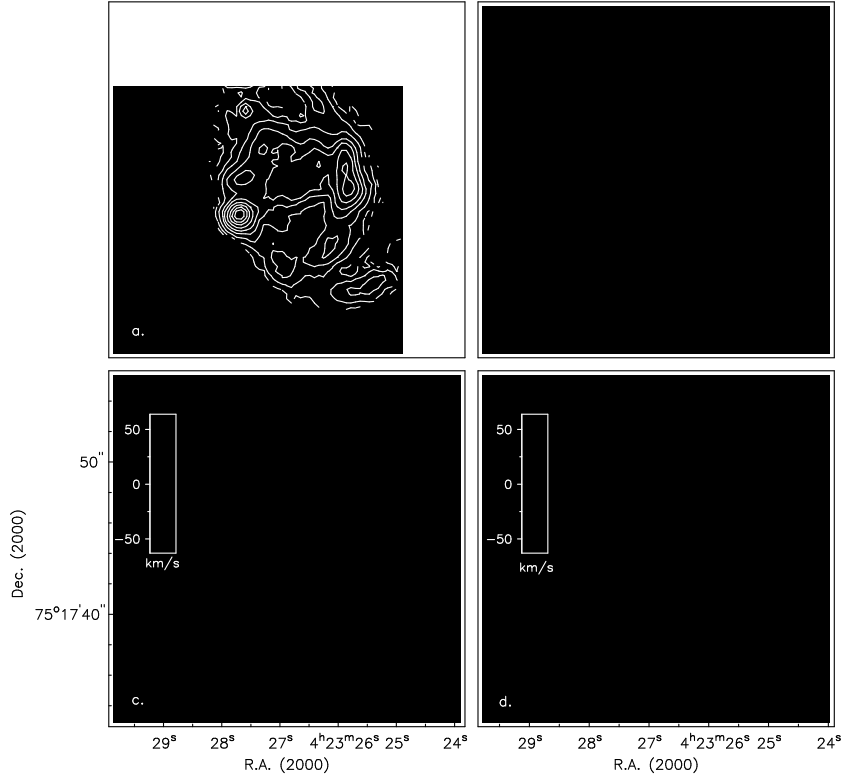}}
\vspace{-0.2cm}
\caption{{\bf (a)} HST--NICMOS H band image of the central disc of NGC~1530, with 
\ha\ contours superposed. {\bf (b)} J--K image of the zone, from P\'erez--Ram\'\i rez 
et al. (2000), clearly showing the interlocking spiral dust lanes around 
the nucleus. {\bf (c)} Residual non--circular velocity map, projected along the 
line of sight, for the circumnuclear disc of NGC~1530, obtained from 
Fabry--P\'erot observations. {\bf (d)} Model predicted velocity field, based on 
constant amplitude flow vectors directed along the paths defined by the
dust lanes in (b). The predictions match very well the observed pattern,
notably within 5\arcsec\ of nucleus, and show the power of the Fabry--P\'erot method for
velocity field diagnostics.}\label{mos_centro}
\end{figure}

	In the inner 5\arcsec\ of our original rotation curve (Zurita et al.
2004) we found a steep gradient which relaxed to a slightly less steep gradient
outside this radius. Assuming this to be caused by a bulge, we used an HST-NICMOS
near-IR image to check this, but as seen in Fig.~\ref{mos_centro}a the 
central spheroidal component is diminutive, and could not cause the steep inner 
gradient. Assuming that we are detecting projected non--circular motion, we
extrapolated the outer part of the rising rotation curve linearly to the origin,
and subtracted this from the observed inner velocity field, yielding a map of 
projected non--circular velocity in the inner disc, as shown in Fig.~\ref{mos_centro}c. This
``yin/yang'' pattern reveals spiral inflow to the nucleus. To show this we took
an image of the zone in J--K from P\'erez-Ram\'\i rez et al. (2000), see Fig.~\ref{mos_centro}b,
which shows interlocking spiral dust lanes. Assuming that these indicate 
inward flow lines we computed the projected velocity we would see, taking for 
simplicity a constant velocity amplitude along the flow. The result, in Fig.~\ref{mos_centro}d
agrees nicely with the observational field in Fig.~\ref{mos_centro}c, except at the north
and south ends of the inner disc, where we know that the \ha\ data do not
allow us to plot the true circular rotation curve, so the residual map in 
Fig.~\ref{mos_centro}c will have systematic errors here. This inward spiralling flow is 
not due to an inner bar. In Zurita et al. (2004, 2004a) we show using unsharp
masking of the HST image that any circumnuclear bar must be less than 0.5\arcsec\
in length, and with negligible dynamical effect over the inner disc. However,
as modelled by Englmeier \& Shlosman (2000) in a galaxy with an inner 
Lindblad resonance at a radius within the length of a major bar, as the dominant
stellar $x_1$ orbits in the bar give way to perpendicular $x_2$ orbits within the ILR 
the gas is pulled round into opposing spiral trajectories, very much as seen 
in Fig.~\ref{mos_centro}, which take it down to within a couple of hundred parsecs 
of the nucleus.
Fabry--P\'erot velocity fields from ionized gas emission give us an excellent
tool for exploring these dynamical effects.

\section{Some general conclusions}
The message of this article is the same for the two different
astrophysical systems considered: galaxy evolution cannot be well understood 
without a knowledge of gas flows and their effects on star formation. For the
Milky Way our reasoning has been indirect. Chemical evidence is used to show
that gas has been accreting to the disc during its whole lifetime, and at rates
which do not appear to have declined during this period. This does not 
contradict the general picture in which the SFR in the universe has been 
declining steadily from $z=1$ to the present epoch, as the overall density has
fallen. We note that the ongoing global disc SFR is of order 2\Msun~ yr\me, while 
during the formation of the initial spherical component (stellar halo) the
rate must have been much higher. This means that a constant, even slowly 
increasing SFR in the discs of galaxies and small groups, regulated by 
inflow and feedback processes, need not contradict inferences about the 
universal SFR from the Madau et al. (1996) plot and its subsequent refinements.
One feature of star formation in galactic discs is its apparently low 
efficiency averaged over time. In a model where the disc mass has grown 
slowly via accretion, this low efficiency is only apparent, an effect of 
time averaging. Any gas which arrives as infalling \hi\ suffers relatively 
rapid conversion to stars, subject to feedback from the massive stars
themselves. This scenario of gas persistently raining down onto galaxy discs,
with the occasional sharper shower, a more massive \hi\ cloud, is finding 
support as \hi\ detection sensitivities increase, and more \hi\ clouds falling
in to other galaxies are detected. And infall can explain rather well what
other mechanisms (major interactions apart) cannot: how the star formation 
rate can increase and decrease simultaneously over the whole, or the major
part of a Milky Way sized disc galaxy. We have considered here the evidence
that this does happen, both for the Milky Way and for galaxies in general.

In the second part of the paper we have shown how gas flow is a key 
parameter determining the star forming pattern in NGC~1530. While shocks,
strong velocity gradients along the line of flow, yield massive stars, 
shear, strong velocity gradients perpendicular to the flow, inhibit massive
star production. NGC~1530 is a specially favourable case for study, for the
essentially geometrical and morphological reasons explained in Sect.~7,
but we should treat it here as an example revealing processes of general
importance. Here we have been able to present a semi--quantitative picture,
but to disentangle fully the effects produced by gas flow on star formation
we will need, as observers, to combine similar information from ionized, neutral
(atomic) and molecular gas. Only then will we have the basis for full 
quantitative understanding, aimed at the major prize: a physical theory of
star formation in different galactic environments.
\vspace{0.2cm}

{\small\em 
The research discussed in this article was supported by
grants AYA2001-0435 (Spanish Ministry of Science and Technology) and 
AYA2004-08251-C02-01 (Spanish Ministry of Education and Science). A. Zurita acknowledges 
support by the Consejer\'\i a de Educaci\'on y Ciencia de la Junta de Andaluc\'\i a, Spain.
The authors thank the organizers of the conference for the invitation to present this
paper in an excellent context, both scientific and social.
}

\end{document}